\documentstyle[12pt,twoside]{article}

\catcode `\@=11
\@addtoreset{equation}{section}
 
\def\theequation{\arabic{section}.\arabic{equation}}
\def\section{\@startsection {section}{1}{\z@}{-3.5ex plus -1ex minus
     -.2ex}{2.3ex plus .2ex}{\normalsize\bf}}
\def\subsection{\@startsection{subsection}{2}{\z@}{-3.25ex plus -1ex minus
 -.2ex}{1.5ex plus .2ex}{\normalsize\bf}}
\def\thebibliography#1{\section*{{
				   \rm REFERENCES}\@mkboth
  {REFERENCES}{REFERENCES}}\list
  {[\arabic{enumi}]}{\settowidth\labelwidth{[#1]}\leftmargin\labelwidth
  \advance\leftmargin\labelsep
  \usecounter{enumi}}
  \def\newblock{\hskip .11em plus .33em minus -.07em}
  \sloppy
  \sfcode`\.=1000\relax}
 
\catcode `\@=12
\title{\vspace*{1.5cm} \normalsize\bf 
DE DONDER--WEYL THEORY AND A HYPERCOMPLEX EXTENSION OF 
QUANTUM MECHANICS TO FIELD THEORY 
}

\author{\vspace{1.5cm} \sc Igor V. Kanatchikov\thanks{e-mail: 
ikanat@ippt.gov.pl, kai@fuw.edu.pl} \vspace{-1.5cm}\\    
\small Laboratory of Analytical Mechanics and Field Theory \vspace{-0.2cm}\\ 
\small Institute of Fundamental Technological Research\vspace{-0.2cm} \\
\small Polish Academy of Sciences\vspace{-0.2cm}\\
\small \'Swi\c etokrzyska 21,  Warszawa  PL-00-049, Poland 
}
\date{\small \it (Submitted May 1998   ------  Accepted September 1998) } 

\begin{document}

\maketitle 

\markboth{\centerline{\small  I.V. KANATCHIKOV}}{ \hspace*{-12.5pt}
\centerline {\small HYPERCOMPLEX EXTENSION OF 
QUANTUM MECHANICS TO FIELD THEORY }}

%newcommands

\vspace*{-96mm}
\hbox to 6.2truein{%\tiny \it 
\footnotesize\it 
to appear in   
\hfil \hbox to 0 truecm{\hss 
\normalsize\rm %Preprint %(May 1998) 
				   }\vspace*{-1mm}}
\hbox to 6.2truein{%\tiny %
\vspace*{-1mm}\footnotesize 
Rep. Math. Phys. 
\hfil 
} 
\hbox to 6.2truein{%\tiny %
\vspace*{-1mm}\footnotesize 
vol. {\bf 42} (1998)   \hfil 
\hbox to 0 truecm{ 
\hss \normalsize hep-th/xxxxxxx} 
}
\vspace*{72mm}

%%%%%%%%%%%%%%%%%%%%%%%%%%%%%%%%%%%%%%%%%%%%%%%%%%%%%%%%%%%%%%%%%%
% ABSTRACT %%%%%%%%%%%%%%%%%%%%%%%%%%%%%%%%%%%%%%%%%%%%%%%%%%%%%%%
%%%%%%%%%%%%%%%%%%%%%%%%%%%%%%%%%%%%%%%%%%%%%%

\begin{flushright} 
\begin{minipage}{5.3in}
{\footnotesize 
A quantization of field theory based on the 
De Donder-Weyl (DW) covariant Hamiltonian 
formulation is discussed. 
A hypercomplex  extension of 
	%the quantum mechanical formalism,    
quantum mechanics,      
in which the space-time Clifford algebra 
    %a hypercomplex algebra 
%plays the role similar to that of 
replaces that of the complex numbers, %% in quantum mechanics, 
appears as a result of quantization of Poisson brackets  
of differential forms 
%which were 
put forward for the 
DW formulation earlier.
The proposed 
 covariant 
 hypercomplex 
%analogue of the 
Schr\"odinger equation 
%, with the wave function argued to be spinor-valued.  
is shown to lead in the classical limit  
to the DW Hamilton-Jacobi equation 
and to obey the Ehrenfest principle in the sense that 
the DW canonical field equations are satisfied for 
the expectation values of properly chosen operators. 
}   
\end{minipage}
\end{flushright} 
%%%%%%%%%%%%%%%%%%%%%%%%%%%%%%%%%%%%%%%%%%%%%%%%%%%%%%%
%%  END ABSTRACT  %%%%%%%%%%%%%%%%%%%%%%%%%%%%%%%%%%%%%
%%%%%%%%%%%%%%%%%%%%%%%%%%%%%%%%%%%%%%%%%%%%%%%%%%%%%%%

\newcommand{\oldtitle}{

\begin{document}
 
\vspace*{2.5cm}
	\begin{center}%%my addition
\noindent
{ \bf DE DONDER--WEYL THEORY AND A HYPERCOMPLEX EXTENSION OF 
QUANTUM MECHANICS TO FIELD THEORY }\vspace{1.3cm}\\
	\end{center}%%my addition
\noindent
\hspace*{1in}
\begin{minipage}{13cm}
Igor V. Kanatchikov$^{1}$  \vspace{0.3cm}\\
 $^{1}$ Laboratory of Analytical Mechanics and Field Theory \\
\makebox[3mm]{ }Institute of Fundamental Technological Research \\
\makebox[3mm]{ }Polish Academy of Sciences \\
\makebox[3mm]{ }\'Swi\c etokrzyska 21, Warszawa PL-00-049, Poland\\
\end{minipage}

\vspace*{0.5cm}

\begin{abstract}
\noindent
A quantization of field theory based on the 
De Donder-Weyl (DW) covariant Hamiltonian 
formulation is discussed. 
A hypercomplex  extension of 
	%the quantum mechanical formalism,    
quantum mechanics,      
in which the space-time Clifford algebra 
    %a hypercomplex algebra 
%plays the role similar to that of 
replaces that of the complex numbers, %% in quantum mechanics, 
appears as a result of quantization of Poisson brackets  
of differential forms 
%which were 
put forward for the 
DW formulation earlier.
The proposed 
 covariant 
 hypercomplex 
%analogue of the 
Schr\"odinger equation 
%, with the wave function argued to be spinor-valued.  
is shown to lead in the classical limit  
to the DW Hamilton-Jacobi equation 
and to obey the Ehrenfest principle in the sense that 
the DW canonical field equations are satisfied for 
the expectation values of properly chosen operators. 
\end{abstract}

} 
%%%%%%%%%%%%%%%%%%%%%%%%%%%%%%%%%%%%%%%%%%%%%%%%%%%%%%%%%%%%%%
%%%%%%%%%%%%%%%%%%%%%%%%%%%%%%%%%%%%%%%%%%%%%%%%%%%%%%%%%%%%%%

%%%%%%%%%%%%%%%%%%%%%%%%%%%%%%%%%%%%%%%%%%%%%%%%%%%%%%%%%%%%%%%%%%%%%%%%%
		%%%%%%%%%%%% File ncom.tex %%%%%%%%%%%%%%%
%%%%%%%%%%%%%%%%%%%%%%%%%%%%%%%%%%%%%%%%%%%%%%%%%%%%%%%%%%%%%%%%%%%%%%%%%

\newcommand{\beq}{\begin{equation}}
\newcommand{\eeq}{\end{equation}}
\newcommand{\beqa}{\begin{eqnarray}}
\newcommand{\eeqa}{\end{eqnarray}}
\newcommand{\nn}{\nonumber}

\newcommand{\half}{\frac{1}{2}}

\newcommand{\xt}{\tilde{X}}

\newcommand{\uind}[2]{^{#1_1 \, ... \, #1_{#2}} }
\newcommand{\lind}[2]{_{#1_1 \, ... \, #1_{#2}} }
\newcommand{\com}[2]{[#1,#2]_{-}} 
\newcommand{\acom}[2]{[#1,#2]_{+}} 
\newcommand{\compm}[2]{[#1,#2]_{\pm}}

\newcommand{\lie}[1]{\pounds_{#1}}
\newcommand{\co}{\circ}
\newcommand{\sgn}[1]{(-1)^{#1}}
\newcommand{\lbr}[2]{ [ \hspace*{-1.5pt} [ #1 , #2 ] \hspace*{-1.5pt} ] }
\newcommand{\lbrpm}[2]{ [ \hspace*{-1.5pt} [ #1 , #2 ] \hspace*{-1.5pt}
 ]_{\pm} }
\newcommand{\lbrp}[2]{ [ \hspace*{-1.5pt} [ #1 , #2 ] \hspace*{-1.5pt} ]_+ }
\newcommand{\lbrm}[2]{ [ \hspace*{-1.5pt} [ #1 , #2 ] \hspace*{-1.5pt} ]_- }
\newcommand{\pbr}[2]{ \{ \hspace*{-2.2pt} [ #1 , #2 ] \hspace*{-2.55pt} \} }
\newcommand{\we}{\wedge}
\newcommand{\dv}{d^V}
\newcommand{\nbrpq}[2]{\nbr{\xxi{#1}{1}}{\xxi{#2}{2}}}
\newcommand{\lieni}[2]{$\pounds$${}_{\stackrel{#1}{X}_{#2}}$  }

\newcommand{\rbox}[2]{\raisebox{#1}{#2}}
\newcommand{\xx}[1]{\raisebox{1pt}{$\stackrel{#1}{X}$}}
\newcommand{\xxi}[2]{\raisebox{1pt}{$\stackrel{#1}{X}$$_{#2}$}}
\newcommand{\ff}[1]{\raisebox{1pt}{$\stackrel{#1}{F}$}}
\newcommand{\dd}[1]{\raisebox{1pt}{$\stackrel{#1}{D}$}}
\newcommand{\nbr}[2]{{\bf[}#1 , #2{\bf ]}}
\newcommand{\der}{\partial}
\newcommand{\oo}{$\Omega$}
\newcommand{\Om}{\Omega}
\newcommand{\om}{\omega}
\newcommand{\eps}{\epsilon}
\newcommand{\si}{\sigma}
\newcommand{\Lm}{\bigwedge^*}

\newcommand{\inn}{\hspace*{2pt}\raisebox{-1pt}{\rule{6pt}{.3pt}\hspace*
{0pt}\rule{.3pt}{8pt}\hspace*{3pt}}}
\newcommand{\sro}{Schr\"{o}dinger\ }
\newcommand{\bm}{\boldmath}
\newcommand{\vol}{\omega}%%{\widetilde{vol}}                                             \newcommand{\dvol}[1]{\der_{#1}\inn \vol}

\newcommand{\bd}{\mbox{\bf d}}
\newcommand{\bder}{\mbox{\bm $\der$}}
\newcommand{\bI}{\mbox{\bm $I$}}

\newcommand{\ga}{\gamma} 
\newcommand{\gmu}{\gamma^\mu}
\newcommand{\gnu}{\gamma^\nu}
\newcommand{\ka}{\kappa}
\newcommand{\hka}{\hbar \kappa}
\newcommand{\al}{\alpha}
\newcommand{\lapl}{\bigtriangleup}
\newcommand{\psib}{\overline{\psi}}
\newcommand{\Psib}{\overline{\Psi}}
\newcommand{\derts}{\stackrel{\leftrightarrow}{\der}}
\newcommand{\what}[1]{\widehat{#1}}

\newcommand{\bx}{{\bf x}}
\newcommand{\bk}{{\bf k}}
\newcommand{\bq}{{\bf q}}

\newcommand{\omk}{\omega_{\bf k}}

%%%%%%%%%%%%%%%%%%%%%%%%%%%%%%%%%%%%%%%%%%%%%%%%%%%%%%%%%%%%%%%%%%%%%%%%%
%%%%%%%%%%%%%%%%%%%%%%%%%%%%%%%%%%%%%%%%%%%%%%%%%%%%%%%%%%%%%%%%%%%%%%%%%
%%%%%%%%%%%%%%%%%%%%%%%%%%%%%%%%%%%%%%%%%%%%%%%%%%%%%%%%%%%%%%%%%%%%%%%%%

\section{\hspace{-4mm}.\hspace{2mm}Introduction }

It is commonly believed in theoretical physics 
that a generalization of the Hamiltonian formalism to field 
theory requires a distinction between the space and 
time variables and implies the treatment of  fields as 
 infinite dimensional mechanical systems. 
However,  another approach is 
%%also 
possible.  It 
treats the space and time coordinates on  equal footing 
(as analogues of a single time parameter in mechanics) 
and does not explicitly refer to 
%%the 
an idea of 
%%the 
a field as a 
mechanical system evolving in time by treating the  field 
rather as a system varying both in space and in time. 
The approach has been known 
as the De Donder--Weyl (DW) theory  
in the calculus of variations  
since the thirties \cite{dedonder} although  
its applications  in physics have been  rather rare.   
For recent discussions of 
%various 
mathematical issues 
 %related to the 
of DW theory and further references see 
\cite{gimm,sardan,ikanat0}. 

Usually the Hamiltonian formalism serves as 
%a starting point 
 a basis 
for the canonical quantization. It is quite natural, therefore, to 
ask whether the DW formulation, viewed as a field theoretic 
generalization of the Hamiltonian formalism in mechanics,    
 can lead to a corresponding quantization 
procedure in field theory. 
	%The purpose of 
In the present paper we discuss an approach to such a quantization 
(for earlier discussions see \cite{ikanat1,qs96}).   
It is our hope that 
the  study of  quantization based on the DW theory can  
contribute to  our 
%current 
understanding of the 
fundamental issues   of quantum field theory 
and to provide us with a new framework of quantization  which could be 
useful in situations where the applicability of the 
conventional canonical quantization in field theory 
can be in doubt. 
% such as the quantization 
%on not globally hyperbolic space-times or 
%the quantization of general relativity.  
Note 
also that the manifest covariance of the approach can make it 
especially appealing in the context of quantization of 
gravity and extended objects. 

\section{\hspace{-4mm}.\hspace{2mm}
De Donder-Weyl theory: a reminder}

Let us recall the essence of the DW formulation.   
Given 
a Lagrangian density 
$L=L(y^a, \der_\mu y^a, x^\nu)$, where 
$\{y^a \}$ are 
field variables, 
$\{ \der_\mu y^a \}$ denote  
their space-time derivatives 
and $\{x^\mu \}$, $\mu=1,...,n,$ are 
space-time coordinates, 
we can define the new set of Hamiltonian-like variables: 
$p_a^\mu:=\der L/ \der (\der_\mu y^a)$,  
called  {\em polymomenta}, 
and  
$H:= \der_\mu y^a p_a^\mu - L$,  
called  the {\em DW Hamiltonian function},  
which allow us to write the Euler-Lagrange field 
equation in an appealing, manifestly covariant 
first order form 
\beq
\der_\mu y^a = \der H / \der p^\mu_a, 
\quad \der_\mu p^\mu_a = - \der H/ \der y^a 
\eeq 
referred to as  the {\em DW Hamiltonian field equations}. 

Similar to the Hamiltonian formulation in mechanics 
an analogue of the Hamilton-Jacobi (HJ) theory can be 
developed 
for the DW Hamiltonian field equations.  
%The corresponding Hamilton-Jacobi equation, to be 
%referred to as the DWHJ equantion, 
It is formulated in terms 
of $n$ 
%Hamilton-Jacobi 
HJ functions on the field configuration space  
$S^\mu=S^\mu(y^a,x^\mu)$ which fulfill the DW HJ  equation
\beq
\der_\mu S^\mu + H(x^\mu, y^a, p_a^\mu = \der S^\mu / \der y^a )=0.   
\eeq 
The quest of 
%%a 
a formulation of a quantum field theory which in the classical limit 
would give rise to the DW HJ equation  has been  one
 of the  motivations of the present study. 

%Note that within the DW Hamiltonian 
%formulation the field is  essentially viewed 
%as a system varying in all space-time directions treated on 
% equal footing. This differs from the conventional Hamiltonian 
%formalism where  the field is treated as a mechanical 
%system of an infinite number of degrees of freedom evolving with time. 

Let us consider an example of interacting scalar 
fields $y^a$ described  by the Lagrangian 
density 
\beq
L= \half \der_\mu y^a \der^\mu y_a - V(y). 
\eeq
Then the polymomenta and the DW Hamiltonian function are given by 
\beq 
p^a_\mu= \der_\mu y^a, \quad 
H= \half p^a_\mu p_a^\mu + V(y), 
\eeq
the DW Hamiltonian field equations take the form 
\beq
\der_\mu y^a =  p_\mu^a,  
\quad \der_\mu p^\mu_a = - \der V/ \der y^a, 
\eeq 
and the DWHJ equation reads 
\beq
\der_\mu S^\mu + \half \der_a S^\mu \der^a S_\mu + V(y) = 0. 
%H(x^\mu, y^a, p_a^\mu = \der S^\mu / \der y^a ). 
\eeq

\section{\hspace{-4mm}.\hspace{2mm} 
Poisson bracket of forms: properties  and the equations of motion} 

To develop an analogue of the canonical quantization procedure 
we  need the Poisson bracket possessing appropriate algebraic properties, 
%%the 
a notion of the canonically conjugate variables, 
and 
%%the 
a representation of the field equations in 
terms of the Poisson bracket. 
%(which/that may allow us to guess 
%the quantum dynamics law). 

In  previous papers \cite{ikanat0,ikanat1,ikanat3} 
we have shown that the proper analogue of the Poisson bracket 
for the DW Hamiltonian formulation can be  defined on horizontal 
differential $p$-forms 
$$\ff{p}:=\frac{1}{(n-p)!}F\uind{\mu}{n-p}(z^M) %(y^a,p_a^\mu,x^\mu) 
\der\lind{\mu}{n-p},
\inn\omega $$ 
$p=0,1,...,(n-1)$,  
which naturally play 
%%the 
a role of dynamical variables. 
The following notations are  used throughout 
\beqa
\omega &:=& dx^1\we ... \we dx^n , \nn \\
\der\lind{\mu}{q} &:= & \der_{\mu_1}\we ... \we \der_{\mu_q} ,  \nn\\
\omega\lind{\mu}{q} & := & \der\lind{\mu}{q}\inn \omega \nn , \nn \\
\{z^M \} & :=&  \{y^a,p_a^\mu,x^\mu \} .
\eeqa 
The sign $\inn$ denotes the inner product of a multivector field 
with a form, such that e.g. $\der_\mu \inn dx^\nu = \delta_\mu^\nu$. 
The same symbol $\der_M$ denotes either the partial derivative with respect 
to the variable $z^M$ or a tangent vector $\frac{\der}{\der z^M}$  
according to  the context.  
For details of the construction of the 
Poisson bracket and its properties we refer to \cite{ikanat0,ikanat3}.  
For us it is most important here 
that the bracket defined on forms leads to several 
generalizations of the  Poisson algebra 
of functions in mechanics, and that it also 
%allows 
enables us to represent the 
equations of motion of dynamical variables in terms of 
the bracket with the DW Hamiltonian function. 

In particular, 
on the class of specific forms, called in \cite{ikanat0,ikanat1} 
Hamiltonian, 
the bracket determines the 
structure of the so-called Gerstenhaber algebra, a specific 
graded generalization of the Poisson algebra. 
By definition, it 
is a graded commutative algebra equipped with a graded 
Lie bracket operation which fulfills  the 
%(left and right) 
graded Leibniz rule with respect to the 
 graded commutative product in the algebra. 
The  grade of 
an element of the algebra with respect 
to the product differs by one from its grade with respect to the 
bracket operation.  

The graded commutative (associative) product 
on Hamiltonian forms is what we called the {\em co-exterior product} and 
denoted $\bullet$.  
It is defined as follows\footnote{Prof. Z. Oziewicz pointed out to 
the author that this product was introduced much earlier 
by Plebanski, see \cite{plebanski}.}: 
\beq
\ff{p}\bullet \ff{q} := *^{-1}(*\ff{p}\we * \ff{q}), 
\eeq 
where $*$ denotes the Hodge duality operator acting on horizontal forms 
and $*^{-1}$ is its inverse. 
As a consequence,    
$\ff{p}\bullet \ff{q}= (-1)^{(n-p)(n-q)}\ff{q}\bullet \ff{p}$,  
deg$(\ff{p}\bullet \ff{q})=p+q-n$, 
%As a result 
and a form of degree $p$ has a grade $(n-p)$ with respect to the 
co-exterior product. 

The bracket operation on Hamiltonian forms 
is graded Lie, with the 
grade of a bracket with a $p$-form being $(n-p-1)$, 
so that  the bracket of a $p$-form 
with a $q$-form is a form of degree $q-(n-p-1)$. The bracket 
also fulfills the graded Leibniz rule with respect to the 
$\bullet$-product
\beq
\pbr{\ff{p}}{\ff{q}\bullet \ff{r}}=
\pbr{\ff{p}}{\ff{q}}\bullet \ff{r} 
+ (-)^{(n-q)(n-p-1)}\ff{q}\bullet\pbr{\ff{p}}{\ff{r}}. 
\eeq
All these properties 
%describe 
characterize 
the space of Hamiltonian 
forms as  a Gersten\-ha\-ber algebra. 

Hamiltonian forms of non-zero degree 
are 
% can be characterized as  
polynomials of 
$(n-1)$-forms $p_a^i\omega_i$ with respect to the $\bullet$-product,  
with  the coefficients being 
 arbitrary functions of the field and 
 space-time variables (cf. eq. (2.4) in \cite{ikanat3}). 
Note that 
the variables $p_a^i\omega_i$ can be viewed as canonically conjugate 
to the  field variables since their Poisson bracket is 
$$
\pbr{p_a^\mu\omega_\mu}{y^b}=\delta^b_a . 
$$ 
In fact, owing to the implicit graded canonical symmetry in the 
theory 
%%(the ``polysymplectomorphisms'') 
there are other canonical pairs of forms of various degrees 
corresponding to the field variables and polymomenta (cf. sect. 4.1). 
The corresponding 
canonical brackets are of particular interest from the point of view of 
the canonical quantization.  

Note, that  a bracket of any two Hamiltonian 
forms can be calculated using the 
canonical brackets and 
the graded Leibniz property of the bracket, 
% Therefore, the algebra 
%of Hamiltonian forms can in principle be constructed axiomatically, 
independently of the construction  in our previous papers 
which uses the notion of the polysymplectic form and the related map 
from forms to multivector fields.  
%based on the postulated canonical brackets 
%%(whatever their motivation cold be)  
%and the algebraic properties of the bracket. 
However, still it is not clear how the co-exterior product, 
the space of Hamiltonian forms, and the canonical brackets could be 
invented or motivated independently of the construction in 
\cite{ikanat0}. % (that it without knowing the answer first). 

The equations of motion  can be written in terms of 
the Poisson bracket of forms.  An analogy with mechanics suggests  
that they are given by the bracket with the DW Hamiltonian function. 
However, the degree counting shows that 
the bracket with $H$ exists  only for Hamiltonian forms of 
degree $(n-1)$: $F:=F^\mu\omega_\mu$. 
For these forms  the equations of motion can be written 
in the form 
\beq
%-\sigma (-1)^{n}
\bd\!\bullet\! F = -\sigma (-1)^{n}\pbr{H}{F} 
+ d^h \!\bullet \! F
\eeq
where $\bd\bullet$ denotes the operation of the ``total 
co-exterior differential'' 
%which is defined as follows
\beq
\bd\!\bullet\! \ff{p}:= \frac{1}{(n-p)!}\der_M F\uind{\mu}{n-p}\der_\mu z^M 
dx^\mu\bullet \der\lind{\mu}{n-p} \omega,  
\eeq 
$d^h$ is a ``horizontal co-exterior differential'': 
$$
d^h\!\bullet \! \ff{p}:= \frac{1}{(n-p)!}\der_\mu F\uind{\mu}{n-p} 
dx^\mu\bullet \der\lind{\mu}{n-p} \omega, 
$$
and $\sigma=+1$ $(-1)$ for the Euclidean (Minkowskian) signature of the 
space-time metric.  
It is evident that  co-exterior differentials identically vanish  
on forms of degree lower than $(n-1)$ sharing  the property 
with the operation of the bracket with $H$. 

Note, that the form of the equations of motion in (3.4) 
is different from that presented in our previous papers 
in which the left hand side has been written in terms of the operator 
$*^{-1}\bd$, where 
%%$*^{-1} $ is the inverse of the Hodge duality 
%%operation on horizontal forms, and 
$\bd$ is a total exterior 
differential defined as follows 
$$
\bd \ff{p}:= \frac{1}{(n-p)!}\der_M F\uind{\mu}{n-p}\der_\mu z^M 
dx^\mu\we \der\lind{\mu}{n-p} \omega. 
$$ 
In  fact, the action of $-\sigma (-1)^{n}\bd\bullet$ 
on $(n-1)$-forms 
coincides with that of 
$*^{-1}\bd$, so that the essence 
of the equations of motion in both 
%forms 
representations remains the same.   
However,  the use of $\bd\bullet$  better conforms with 
the natural product operation $\bullet $  
%on the space 
of Hamiltonian forms and 
also 
with the fact that the bracket with $H$ exists only for forms of 
degree $\geq (n-1)$.  

Note also, that 
the 
Poisson bracket formulation  of the equations of motion can be extended 
(in a weaker sense) to 
arbitrary horizontal forms. For this purpose one have to make 
sense of the bracket with the DW Hamiltonian $n$-form $H\omega$. 
The result is that the bracket with $H\omega$ 
%generates  the  
corresponds to the 
total exterior differential of a form \cite{ikanat0}.

\section{\hspace{-4mm}.\hspace{2mm}
Elements of the canonical quantization} 

\subsection{\hspace{-5mm}.\hspace{2mm}
%The canonical brackets and their quantization} 
Quantization of the canonical brackets} 

The problem of  quantization of the Gerstenhaber algebra of 
Hamiltonian forms is by itself, 
independently of its application to field theory,  
an interesting mathematical problem, 
which could be approached 
%%from the point of view of 
 by 
different mathematical 
techniques of quantization, such as a deformation quantization 
or a geometric quantization. 
However, in this paper  we shall follow a more naive approach 
based on  extending the rules of  the canonical quantization 
to the present framework. 

Let us recall that in quantum mechanics it is sufficient to quantize 
only a small 
%subalgebra 
part 
of the Poisson algebra
%%, that 
	%the center 
given by the canonical brackets. 
Moreover, it is known to be 
impossible to quantize the whole  Poisson algebra due to 
the limits imposed by the Groenewold-van Hove theorem 
(see e.g. \cite{emch}).    
Therefore, to begin with let us confine ourselves 
to an appropriate  
small subalgebra in the algebra of Hamiltonian forms. 

From the properties of the graded Poisson bracket 
discussed in the previous section 
it follows that 
%$(n-1)$-forms and 
the subspace of $(n-1)$-forms and $0$-forms  
constitutes  a Lie subalgebra in the Gerstenhaber 
algebra of Hamiltonian forms. 
Let us quantize the canonical brackets in this 
subalgebra. Nonvanishing brackets are given by \cite{ikanat0}
 $$\pbr{p_a^\mu\omega_\mu}{y^b}
= 
\delta^b_a , \quad 
\pbr{p_a^\mu\omega_\mu}{y^b\omega_\nu}
=
\delta^b_a\omega_\nu, \quad 
\pbr{p_a^\mu}{y^b\omega_\nu}
= 
\delta^b_a\delta^\mu_\nu .  
\refstepcounter{equation} 
\eqno {  (\theequation a,b,c) } 
$$
%\pbr{p_a^\mu}{y^b\omega_\nu}&=&\delta^b_a\delta^\mu_\nu \nn \\
As usual, we associate 
Poisson brackets to commutators divided by $i \hbar$ and 
find the operator realizations of the quantities involved 
on an appropriate Hilbert space.  
In the Schr\"odinger 
$y$-representation from quantization of 
(4.1a) it follows that the operator corresponding to the $(n-1)$-form 
$p_a^i\omega_i$ can be represented by the partial derivative with 
respect to the field variables: 
\beq
\widehat{p_a^\mu\omega_\mu}= i \hbar \der_a. 
\eeq
Quantization of the bracket in 
(4.1b) does not add anything new. 
However, quantization of (4.1c) is nontrivial. 

Let us  write  $\hat{p}{}^\mu_a$ in the form  
$i\hbar  \hat{p}{}^\mu\der_a$, where the operator  
$\hat{p}{}^\mu$ has to be found. Then, from the commutator 
corresponding to (4.1c) we obtain 
\beqa
[\hat{p}{}^\mu_a, \widehat{y^b \omega}_\nu]
&=& i\hbar \hat{p}{}^\mu \der_a \circ \widehat{y^b \omega_\nu}
- %%(-)^{(n-0-1)(n-(n-1)-1)} 
\widehat{y^b \omega_\nu} \circ
i\hbar \hat{p}{}^\mu\der_a 
\nn \\
&=& i\hbar \delta _a^b \hat{p}{}^\mu \circ \widehat{ \omega}_\nu
+\hat{p}{}^\mu \circ \widehat{ \omega}_\nu\hat{y}{}^b i\hbar \der_a
- \widehat{ \omega}_\nu\circ\hat{p}{}^\mu \hat{y}{}^b i\hbar \der_a , 
%= i\hbar \delta^\mu_\nu  \delta^b_a , 
\eeqa
where $\circ $ denotes 
a composition law of operators which implies 
some  not known in advance 
multiplication law of 
horizontal operators 
$\widehat{\omega}_i$ and $\hat{p}{}^i$. 
%% which is not known in advance. 
The right hand side 
of (4.3) will be equal to $i\hbar \delta^\mu_\nu  \delta^b_a$,  
as it is required by (4.1c),  if the following two conditions 
are fulfilled: 
\beq
\hat{p}{}^\mu \circ \widehat{ \omega}_\nu = \delta^\mu_\nu, 
\quad 
\hat{p}{}^\mu \circ \widehat{ \omega}_\nu
- \widehat{ \omega}_\nu\circ\hat{p}{}^\mu = 0.  
\eeq 
%These conditions tell us that 
Hence, the composition law 
$\circ $ is a symmetric operation, 
i.e. 
$\hat{p}{}^\mu \circ \widehat{ \omega}_\nu
=\half (\hat{p}{}^\mu \circ \widehat{ \omega}_\nu
+ \widehat{ \omega}_\nu \circ \hat{p}{}^\mu )$. 
%motivations: 
%geom quant
%gradings
%dirac vs kahler vs grassman
These properties   can be satisfied quite naturally 
by the hypercomplex imaginary units of the Clifford 
algebra of the space-time.  
 %%over which the theory we are quantizing is constructed. 
These hypercomplex imaginary 
units $\ga_\mu$ 
(which in four-dimensional Minkowski space-time 
can be represented by  Dirac matrices) 
are defined by the relation $\ga_\mu\ga_\nu+\ga_\nu\ga_\mu=\eta_{\mu\nu}$, 
where $\eta_{\mu\nu}$   is the space-time metric tensor. 
Then the operators above can be realized as follows
\beq
\hat{p}{}^\nu = - \kappa \ga^\nu ,
\quad 
\widehat{ \omega}_\nu = - \kappa^{-1} \ga_\nu , 
\eeq
where the quantity $\kappa$ of  the dimension 
[{\em length}${}^{-(n-1)}$] appears  in order to account for  
the physical dimensions of $p^\nu$ and $\omega_\nu$. 
Since $\omega_\nu$ is essentially an infinitesimal volume 
element the absolute value of $\kappa^{-1}$ can be expected 
to be very small. As a result, the theory under consideration 
requires  the introduction of 
a certain analogue of the fundamental length from the 
elementary requirement of matching of the dimensions. 

Note that the realization (4.5) in terms of Dirac matrices 
is not uniquely determined by (4.4). 
%Although 
In sect. 5.3 we show that this 
choice is consistent with  the Ehrenfest theorem.   
%%%%%%%%%%%%%%%%%%%%%%%%%%%%%%%%%%%%%%%%%%%%%%%%%%%%%%%%%%%%
Still,  an open question to be investigated is 
whether or not other hypercomplex systems 
can be useful for the realization of 
the commutation relations following from quantization of 
the Poisson brackets of forms. 
%(4.4) is an open question to be investigated.  

\subsection{\hspace{-5mm}.\hspace{2mm} 
DW Hamiltonian operator}

In order to quantize a simple field theoretic model 
given by (2.3) we have to construct the operator corresponding 
to the DW Hamiltonian in (2.4). 
{ Note that we cannot just naively multiply operators. For example, 
from (4.4) 
for the operator $\widehat{p_a^\mu\omega}_\mu$ one would 
obtain 
$%\widehat{p_a^\mu\omega}_\mu= 
\hat{p}{}^\mu_a  \cdot 
\widehat{\omega}_\mu 
= -i n \hbar \der_a$, whereas the correct answer consistent with the 
bracket in (4.1a) is $\widehat{p_a^\mu\omega}_\mu= i\hbar \der_a$. } 
To find the operator of $p^\mu_a p^a_\mu$ 
in (2.4) let us quantize the 
bracket 
\beq
\pbr{\frac{1}{2} p^\mu_a p^a_\mu}{y^b \omega_\nu}= p^b_\nu  . 
\eeq

From the corresponding commutator 
\beq
[ \frac{1}{2} \widehat{p^\mu_a p^a_\mu}, \widehat{y^b \omega_\nu} ]
= \frac{1}{2}( \widehat{p^\mu_a p^a_\mu}
\circ \widehat{y^b \omega_\nu} 
-\widehat{y^b \omega_\nu} \circ
\widehat{p^\mu_a p^a_\mu})
=i\hbar \widehat{p}{}^b_\nu  
\eeq
using the  representations of $\widehat{\omega}_\mu$ 
and $\hat{p}{}^\mu_a$,  and the commutator 
$[\lapl, y^a]= 2\delta^a_b \der^b$, 
where $\lapl:= \der_a \der^a$ 
is the Laplacian operator in the field space,  
we obtain 
$$\widehat{p^\mu_a p^a_\mu}= -\hbar^2 \kappa^2 \lapl .$$ 
Hence  the DW Hamiltonian operator for the system of 
interacting scalar fields takes the form
\beq
\widehat{H} = -\half \hbar^2 \kappa^2 \lapl + V(y) .
\eeq
For a free scalar field $V(y)=\half m^2 y^2/\hbar{}^2$,  
%so that  
and the above  expression is similar to the Hamiltonian of the 
harmonic oscillator in the field space. 

\section{\hspace{-4mm}.\hspace{2mm}
Generalized Schr\"odinger equation} 

Our next step is to formulate a dynamical law. An analogue 
of the Schr\"odinger equation here has to fulfill the following 
natural requirements: 
\begin{itemize}
\item  
the familiar quantum mechanical Schr\"odinger equation 
should be reproduced 
if the number of space-time dimensions $n=1$; 
\item  
the DW HJ equation should arise  
in the classical limit; 
\item 
the classical field equations in the DW canonical form 
should be fulfilled for the expectation values of the 
corresponding operators. 
\end{itemize}
We also imply that basic principles of quantum theory such 
as the superposition principle and the probabilistic interpretation 
should be inbuilt in the 
desired 
%sought-for 
generalization. 
Additional hint comes from the bracket form of the equations 
of motion (3.4) and 
an analogy with quantum mechanics. They 
suggest that the sought-for Schr\"odinger equation 
%is of the form 
has the symbolic form 
$ \hat{\i } \hat{d } \Psi \sim \what{H} \Psi $, 
where $\hat{\i }$ and $\hat{d }$ denote appropriate analogues 
of the imaginary unit and the exterior differentiation respectively.

The above considerations have led us to the following 
generalization of the Schr\"odin\-ger equation 
\beq
\label{seqcl}
i \hbar \kappa \gamma^\mu \der_\mu \Psi = \what{H} \Psi, 
\eeq
where $\widehat{H}$ is the operator corresponding to the 
DW Hamiltonian function, the constant $\kappa$ 
of the dimension [{\em length}]$^{-(n-1)}$ appears again on dimensional 
grounds, and $\Psi=\Psi(y^a,x^\mu)$ is a wave function over the 
configuration space of the field and space-time variables. 
Equation (5.1) leaves 
us with 
two options 
%% for  
as to 
the nature of the 
wave function $\Psi$. 
%It 
The latter can be either a 
%general 
hypercomplex number 
\beq
\Psi = \psi I + \psi_\mu \ga^\mu + 
\psi_{\mu \nu} \ga^{\mu\nu} + 
\psi_{\mu_1 ... \mu_n} \gamma^{\mu_1 ... \mu_n} , 
\eeq 
where $\gamma^{\mu_1 ... \mu_p}:= \ga^{[\mu_1} ... \ga^{\mu_p]}$, 
or a Dirac spinor (the choice of the  Dirac spinors is 
%%due to 
based on the 
fact that they exist in arbitrary space-time dimensions and 
signatures) 
which 
%%in fact 
actually can be understood as an element of a  minimal left ideal in the 
Clifford algebra 
%whose general element has the form (5.2) 
 \cite{hestenes,dk}.  The choice in favor of spinors is made in 
sect. 5.2 on the basis of the consideration of the scalar products. 

%\\ Recall ... ideal, reference spinor.........
%\\ Clifford vs complex numbers 
%\\  extension to other hypercomplex systems is trivial 

\subsection{\hspace{-5mm}.\hspace{2mm}
Quasiclassical limit and  DW HJ equation} 

Let us show that (5.1) leads to the DW HJ equation in the quasiclassical 
limit. It is natural to consider the following generalization 
of the quasi-classical ansatz 
\beq
\label{qucl}
\Psi = R \, \exp (iS^\mu \ga_\mu / \hka ), 
\eeq
where $R$ and $S^\mu$ are functions of both the field and space-time 
variables. The exponent in (5.3) is understood as a series 
expansion so that one has the analogue of the Euler formula 
\beq
\exp(i \al S^\mu \ga_\mu)
= \cos \, \al |S| + i \ga_\mu \,\frac{S^\mu}{|S|}  \sin \, \al |S|, 
\eeq
where  
%$|S|:=\mbox{\rm sign}(S^\mu S_\mu)\sqrt{|S^\mu S_\mu|}$ 
$|S|:= \sqrt{S^\mu S_\mu}$ can be both real and imaginary, 
and  $\al := 1/\hka $. 
Thus the 
nonvanishing components of the 
quasiclassical wave function (5.3) are as follows
\beq
\psi = R \, \cos \, \al |S|,   
\quad   
\psi^\mu = i R \, \frac{S^\mu}{|S|} \sin \, \al |S| . 
\eeq 

The wave function of the  form 
$$\Psi=\psi + \psi_\mu \ga^\mu $$  
is sufficient to close the system of equations 
which follows from (5.1). Indeed, in this case  (5.1) reduces to 
\beqa
 i\hbar \kappa \der_\mu \psi^\mu &=& \what{H} \psi   , \\
 i\hbar \kappa \der_\mu \psi &=& \what{H}  \psi_\mu , 
\eeqa
and the remaining equation $\der_{[\mu}\psi_{\nu ]}=0,$
which follows from the 
$\ga^{\mu \nu}$-component, 
%%: $$\der_{[\mu}\psi_{\nu ]}=0,$$  
%just 
is equivalent to 
the integrability condition of (5.7) 
%%if  $\der_\mu \what{H} = 0$. 
if $\what{H}$ is assumed to be independent of $x$-s. 

Now, let us substitute (5.5) to (5.6) and (5.7) 
with the DW Hamiltonian operator 
given by (4.8) and collect together the terms appearing with the 
{\it cos} and {\it sin}  functions respectively. 
Then from (5.6) we obtain  
\beqa
\label{cos-1}
R \, \frac{S^\mu \der_\mu |S|}{|S|}
&=& -\half ( R\, \der_a|S| \der_a |S| 
- \al^{-2} \lapl R 
+ \frac{m^2}{\hbar^2} y^2 R )
\\
\label{sin-1}
\hspace*{-15pt}
\frac{R}{|S|}\der_\mu S^\mu 
&=&
-\half (R\lapl |S| +2 \der_a R \ \der_a |S|)
+ \frac{R S^\mu \der_\mu |S| - |S| S^\mu \der_\mu R}{|S|^2} , 
\eeqa
and (5.7) yields 
\beqa
\label{cos-2}
\der_\mu R &=& -\half \frac{S^\mu}{|S|}\der_a R \ \der_a |S| - 
\half R \frac{S^\mu}{|S|} \lapl |S| 
- \half R \ \der_a  \frac{S^\mu}{|S|} \ \der_a |S|,    
\\
\label{sin-2}
R\ \der_\mu |S| &=& -\half \frac{m^2}{\hbar^2} y^2 R \ \frac{S^\mu}{|S|} 
+ \half \al^{-2}  \frac{S^\mu}{|S|} \lapl R
+ \half \al^{-2} \der_a R \ \der_a \frac{S^\mu}{|S|} 
\nn \\
&&+ \half \al^{-2} R \lapl \frac{S^\mu}{|S|} 
+ \half R \frac{S^\mu}{|S|} \der_a |S| \  \der^a |S|  . 
\eeqa
By contracting (\ref{cos-2})  with $2R \frac{S^\mu}{|S|}$ 
and using (\ref{sin-1}) we obtain 
\beq
\label{ura}
\der_\mu S^\mu = \frac{S^\mu}{|S|} \der_\mu |S| . 
\eeq
Similarly,  eq. (\ref{cos-1})  and eq. (\ref{sin-2}) contracted with
$\frac{S^\mu}{|S|} $  yield 
\beq
\label{urb}
\der_a S^\mu \der^a S_\mu=\der_a|S| \der^a |S|.  
\eeq
With the aid of  (5.12) and (5.13) equation  (5.8) can be written 
in the form 
\beqa
\label{dwhjq}
\der_\mu S^\mu = -\half \der_a S^\mu \der_a S_\mu 
- \half \frac{m^2}{\hbar^2} y^2 
+\half \hbar^2 \kappa^2 \frac{\lapl R }{R}. 
\eeqa 
Obviously, in the classical limit $\hbar \rightarrow 0$  
(5.14) reduces to the DW HJ equation (2.6). However, 
%in parallel with 
besides (5.14) the quasiclassical ansatz (5.3) leads 
to  two supplementary conditions (5.12) and (5.13) on HJ functions 
$S^\mu$. These conditions are just  trivial identities 
at $n=1$. At $n>1$ they represent a kind of duality between 
the field theoretical Hamilton-Jacobi formulation in terms 
of $n$ functions $S^\mu$ and the mechanical-like Hamilton-Jacobi 
equation (in the space  of field variables) for the 
eikonal function $|S|$,  with the analogue of the time derivative   given 
by the directional derivative $\frac{S^\mu}{|S|} \der_\mu$. 
On the one hand, 
a possible speculation could be that this is just another 
manifestation 
of a quantum duality between the particle and the field (wave) 
aspects of a quantum field. On the other hand the 
appearance of the supplementary conditions alien to the DW HJ theory 
can be related to the fact that the ansatz (5.3) does not represent 
the most general hypercomplex number: instead of $2^n$ components 
we have in (5.3) only $(n+1)$ independent functions. The most 
general ansatz would be 
$$%\beq
\Psi = R \exp \{i (S^\mu\ga_\mu + S^{\mu\nu}\ga_{\mu\nu} 
+...+ S\uind{\mu}{n}\ga\lind{\mu}{n})/ \hbar \kappa\} . 
$$%\eeq
However, its substitution to (5.1)  leads to cumbersome expressions 
which 
%%I  or we ???  
we have been unable to analyze and, moreover,    
it is not clear  whether 
	%a possibility of the interpretation of 
the antisymmetric quantities 
$S^{\mu\nu},S^{\mu\nu\alpha}\,$ etc. 
can be interpreted within some generalized (maybe  Lepagean?)  
Hamilton-Jacobi theory for fields.
%% is not evident. 

Note, that all the above conclusions  can be extended to the 
case when the wave function in (5.1) is a spinor. 
%%To this end 
For this purpose the quasiclassical 
ansatz for the spinor wave function can be taken in the form 
\beq
%\label{qucl}
\Psi = R \, \exp (iS^\mu \ga_\mu / \hka ) \eta  , 
\eeq
where $\eta $ denotes a constant reference spinor, e.g. 
$\eta=||1,0,...,0||^T$, which allows us to convert a Clifford 
number to an element of an ideal of the Clifford algebra, 
i.e. to a spinor.  The same extends to the presented 
above  more general ansatz.  

\subsection{\hspace{-5mm}.\hspace{2mm}
Scalar products: hypercomplex vs. spinor wave functions}

Let us return now to the issue of the probabilistic interpretation 
of the wave function. Note first that 
if we restrict ourselves to the hypercomplex 
wave functions of the form $\Psi=\psi+\psi_\mu \gamma^\mu$,  
then equations  (5.6), (5.7) and their complex 
conjugates lead to the  conservation law 
\beq
\der_\mu \int dy\, [\psib \psi^\mu + \psi \psib{}^\mu  ] = 0 
\eeq
under the assumption that the wave function sufficiently 
rapidly decays at $|y|\rightarrow \infty$ and that the 
operator of the DW Hamiltonian is Hermitian with respect to 
the $L^2$ scalar product of functions in 
%%%the 
$y$-space.  
{}From (5.16) it follows that 
the spatial integral over a space-like hypersurface $\Sigma$
\beq
\int_\Sigma \omega_\mu \int dy [\psib \psi^\mu + \psi \psib{}^\mu  ]
\eeq
is preserved in time 
(or, equivalently,  does not depend on the variation of the 
hypersurface $\Sigma$) and, therefore,  
could be viewed as a norm of the hypercomplex wave function. 
As this norm  involves the integration over a space-like hypersurface 
it could be useful for the calculation of the expectation 
values 
of global observables. 
However, its significant drawback is that 
it is not necessarily positive definite as 
a consequence of the  similarity  of (5.17) 
with the scalar product in the Klein-Gordon theory.  
The similarity  is evident from (5.7) which essentially states 
that $\psi_\mu \sim i \der_\mu\psi $. 

As a matter of fact, for the purposes of the present theory we 
need rather  a scalar product 
for the calculation of the expectation values of 
 operators representing local quantities. 
This  scalar product should be scalar (to not change 
the tensor behavior of operators under averaging) 
and 
involve 
only the integration over the field 
space dimensions (to keep the local character of the 
quantities under averaging). For the hypercomplex 
wave function of the type $\Psi=\psi+\psi_\mu \gamma^\mu$ 
the scalar product could be chosen in 
the form 
$$\int dy [ \psib \psi  + \psib{}^\mu \psi_\mu]$$ 
which is, however, not positive definite in general. 
In fact, 
non-existence of the appropriate scalar product 
for wave functions taking values in algebras different from 
the real, complex, quaternion and octonion numbers follows 
from the natural axioms 
ensuring the availability of the probabilistic interpretation 
and general algebraic considerations  
(see e.g. \cite{adler}). Moreover, our attempts to use the 
just mentioned scalar product  in order to obtain an analogue of 
the Ehrenfest theorem have failed.  

To avoid, at least partially, the difficulties above, 
we assume that the wave function 
in (5.1) is a Dirac spinor. Indeed, in this case 
the analogue of the global scalar product (5.17)
\beq
 \int_\Sigma \omega_\mu \int dy \Psib \ga^\mu \Psi ,  
\eeq
where $\overline{\Psi}$ denotes the Dirac conjugate of $\Psi$,  
{\em is} positive definite. We 
also can write 
a scalar product 
for the averaging of local quantities 
%\beq \int dy \Psib \Psi \eeq
\beq
<\Phi|\Psi> := \int dy \, \overline{\Phi}{\Psi} .
\eeq
Then the expectation values of 
operators $\what{O}$ are  calculated 
according to the formula
\beq
<\what{O}> := \int dy \, \overline{\Psi}\what{O}{\Psi}.  
\eeq 
However, since the scalar product $\overline{\Psi}\Psi$ 
is not positive definite  the validity of the  averaging 
in (5.20) is 
%open to question. 
questionable.   
Moreover, from the generalized Schr\"odinger 
equation (5.1)  written in the form 
\beq
i\hbar\kappa \der_\mu \Psi = 
- i\hbar\kappa \gamma_{\mu \nu}\der^\nu \Psi   + \gamma_\mu \what{H} \Psi 
\eeq
and its conjugate  
% accounting for the Hermicity of $\what{H}$ 
we  derive
\beq
\der_\mu \int  dy \,\overline{\Psi}{\Psi} = 
-i \int  dy \, [ \overline{\Psi}\gamma_{\mu \nu}\der^\nu{\Psi} 
- \der^\nu \overline{\Psi}\gamma_{\mu \nu}{\Psi}] \neq 0 , 
\eeq
so that the scalar product in (5.19) is space and time dependent, 
the property which makes it unsatisfactory analogue of the scalar product 
of wave functions in quantum mechanics.  
Nevertheless, we show in what follows 
that the use of the formula (5.20)
for the expectation values  allows us to 
obtain an
analogue of the Ehrenfest theorem.

\subsection{\hspace{-5mm}.\hspace{2mm}
The  Ehrenfest theorem} 

Let us consider  the evolution of the expectation values 
of operators calculated according to (5.20). 
%In this section we  are particularly interested in establishing 
%a field theoretic counterpart  of the Ehrenfest theorem 
%known in quantum mechanics. 
Using the generalized Schr\"odinger equation  (5.1) 
for the evolution of the expectation value of the polymomentum 
operator 
$\what{p}{}_a^\mu=- i \hbar \kappa \ga^\mu \der_a$ 
we obtain 
\beqa
\der_\mu <\what{p}{}_a^\mu> 
&=& - \der_\mu \int dy \overline{\Psi}
 i  \hbar \kappa \ga^\mu \der_a{\Psi} \nn\\
&=& -i \hbar \kappa\int dy [ \der_\mu  \overline{\Psi} 
 \ga^\mu \der_a{\Psi} 
+ \overline{\Psi}  \der_a \ga^\mu \der_\mu {\Psi}]  \nn\\
&=&   
\int dy [ (\what{H} \overline{\Psi}) \der_a\Psi 
- \overline{\Psi} \der_a( \what{H}{\Psi}) ] \nn \\ 
&=&- \int dy \overline{\Psi}  (\der_a \what{H}) \Psi \nn \\
&=&- <\der_a \what{H}> ,   
\eeqa
where the Hermicity of $\what{H}$ with respect to the 
scalar product of functions in $y$-space is used.  
Similarly, 
 \beqa
\der_\mu<\widehat{y_a \omega^\mu}> 
&=&- \kappa^{-1} \der_\mu \int dy \overline{\Psi}\ga^\mu y_a  \Psi \nn \\
&=& \frac{i}{\hbar\kappa^2}\int dy [ (\what{H}\overline{\Psi})   y_a  \Psi 
- \overline{\Psi} y_a  (\what{H} \Psi)]  \nn \\
&=&i \hbar \int dy \overline{\Psi}\der_a \Psi \nn \\
&=& <\what{p^\mu_a \omega_\mu}>  . 
\eeqa 
By comparing these results with the DW Hamiltonian equations 
(2.1) we  conclude that the 
latter are fulfilled 
in 
%on ?  
average as  
a consequence of 
(i) the generalized Schr\"odinger equation (5.1), 
(ii) the 
rules of quantization leading to the realization (4.5) of the 
operators, and 
(iii) the prescription (5.19) for the averaging of the 
operators representing local dynamical variables.  

Thus we have arrived at  the field 
theoretic counterpart (within the approach under discussion) of the 
Ehrenfest theorem known in quantum mechanics.     
Its validity 
could be 
%%viewed 
seen as a justification and  consistency check 
of the whole approach of the present paper.  
However, the situation is not that perfect 
because  
%%as 
the norm used for  the calculation of 
the expectation values is neither  positive definite nor  
constant over the space-time. This 
brings about 
   %%results in 
potential 
problems with the probabilistic interpretation and points 
to the need 
for   improvement of  the current formulation 
or for a better understanding  of its 
physical content. 
%Note also, that  
Moreover, the analogue of the Ehrenfest theorem  can be obtained 
only for specially chosen  operators and the principle behind this 
choice   
%selection 
is not clear. 
%%%%%%%%%%%%%%%%%%%%%

Note also, that the presented  proof of the Ehrenfest theorem 
is not sensible 
to the identification of $\ga$-s  with the Dirac matrices. 
In principle, the use of other hypercomplex units  
(which appear in various  first-order relativistic wave equations)    
%such as e.g. the Duffin-Kemmer matrices, 
in the realization of operators 
$\what{p}{}_a^\mu$ and $\what{\omega}{}_\mu$ 
and in the generalized Schr\"odinger equation  
can also be consistent with the Ehrenfest theorem, 
but we lack  an appropriate interpretation of this 
observation.  

%>> not for any operators p and y!!!!!>>Comment!

\section{\hspace{-4mm}.\hspace{2mm}
Conclusion }
The De Donder-Weyl formulation provides us with the alternative 
covariant canonical framework 
	%%to built up
for quantization of field theory. On the classical level 
it possesses  the analogues of the appropriate geometric and 
algebraic structures, 
such as the Poisson bracket with corresponding  Lie and Poisson 
algebraic 
properties, the notion of the canonically conjugate variables, 
and the Poisson bracket formulation of the equations of motion. 
%In doing so 
Within the DW formulation  field theory 
%appears 
is treated essentially as 
a generalized 
multi-parameter, or ``multi-time''  generalized Hamiltonian system 
with  the space-time variables 
%%generalizing 
entering on equal footing as generalizations of  
the time parameter  
in mechanics.  
The configuration space is a finite dimensional 
bundle  of field variables over the  space-time, of which the 
field configurations are the sections,  instead of 
the usual infinite dimensional space of the field configurations 
on a hypersurface of the constant time.   
The  analogue of the canonical formalism for the 
DW formulation \cite{ikanat0} 
arises  
as a graded version of the canonical 
formalism in mechanics, with the role of dynamical 
variables played by differential forms. 

Quantization of the canonical brackets 
leads us to a hypercomplex extension of 
the quantum mechanical formalism, 
with the usual complex quantum mechanics recovered in the limiting 
case of 
a one-dimensional ``field theory'', that is in mechanics. 
In higher dimensions 
 %%%a hypercomplex algebra, 
 %the Clifford (=hypercomplex) 
the Clifford 
algebra of the corresponding space-time 
manifold 
 plays the central  role. Namely,  in the Schr\"odinger picture 
considered here 
the quantum operators are realized as the differential operators 
with hypercomplex coefficients, and the wave functions take values 
in the 
	spinor space, which is known to be 
the minimal left ideal in  the 
Clifford algebra \cite{dk}. 
The generalized Schr\"odinger equation 
%which is 
formulated  
 in sect. 5 can  also be viewed as a 
%``multi-time''  
multi-parameter hypercomplex 
extension of the quantum mechanical Schr\"odinger equation: 
the right hand side of the latter, $i\der_t$, is generalized to 
the Dirac operator $i\gamma^\mu\der_\mu$. 
This equation, with some reservations, 
is  shown to lead  in the classical limit 
to the field theoretic DW Hamilton-Jacobi equation 
 and to give rise 
to the analogue of the Ehrenfest theorem for the evolution of the 
expectation values of operators corresponding to field variables 
and polymomenta. However, a potential problem with the proposed  generalized 
Schr\"odinger equation is that the scalar product involved in the proof 
of the Ehrenfest theorem is 
not positive definite  
%space-time translation invariant. 
and not constant over the space-time.  
%% COMMENT on other hypercomplex systems!!!!

Although we have entertained  
%have adhered to ... 
here a point of view that the space-time 
Clifford algebras play the central role, 
%many 
some of the results, 
except the derivation in sect. 5.1 of the DW HJ equation 
in the quasiclassical limit 
and quantization of the canonical brackets in sect. 4.1,  
seem to hold true if 
%straightforwardly extend to the case when 
 $\ga^\mu$-s 
are generating elements of other hypercomplex systems 
used in the first order relativistic wave equations,  
such as the Duffin-Kemmer ring. 
How the corresponding  non-Clifford hypercomplex 
extensions of quantum mechanics 
can be reconciled with quantization based on the DW theory,  
whether they follow from quantization 
similar to that in sect. 4.1, 
and which extension (Clifford or non-Clifford) 
%corresponds to the physical reality   
can be suitable in physics 
%adequate to the physical reality 
are the questions 
 we hope to address in our 
further research.

Though the proposed quantization scheme reproduces  essential formal 
ingredients of  quantum theory, 
the prospects of its physical applications 
%still 
remain obscure. 
The obstacle is that the conceptual framework of the present 
approach is  different from the usual  one, 
that makes the translation to the 
conventional language of quantum field theory a difficult task. 

A possible link could be established with the functional Schr\"odinger 
picture in quantum field theory \cite{hatfield}. 
%\cite{mansfield,hatfield,jackiw}. 
On the one hand, the Schr\"odinger wave functional $\Psi([y(\bx)],t)$ 
is a probability amplitude of 
finding the field in the configuration $y(\bx)$ on the hypersurface 
with the time label $t$. On the other hand, 
it is natural to interpret 
 % a natural interpretation of 
our wave function $\Psi(\bx,t,y)$ 
 %would be, roughly speaking, 
%as, roughly speaking, 
as a probability amplitude 
of finding the  field value $y$ in the space-time point $(\bx,t)$. 
Hence, the Schr\"odinger wave functional could appear as a kind of 
composition of  amplitudes given by our wave functions taken at all 
points $\bx$ of the space. 
        A more technical discussion of this issue 
in \cite{qs96} points to  a relation between both 
at least in the ultra-local 
approximation of vanishing wave vectors. However,   
beyond this unphysical approximation the relation 
remains conjectural and requires further study. 
 %although the problem of the construction of the 
 %Schr\"odinger functional  from the wave function $\Psi(\bx,t,y)$ 
 %remains unsolved. 

\noindent 

{\sl Note  
%to be 
added:}  
%in proof:} 
In the recent preprint \cite{navarro} 
M. Navarro considered
 an approach to quantization in field 
theory which is similar to the approach of the present paper.

\bigskip
%\pagebreak 

\noindent 
{\bf Acknowledgments }
\medskip 

\noindent 
The work has been partially supported by the 
Stefan Banach Inter\-national Mathematical Center to which I express 
my gratitude. I wish to thank 
M. Flato, 
Z. Oziewicz, 
J. S\l awianowski 
and C. Lopez Lacasta  
for discussions, 
J. Stasheff for his useful comments on the author's paper \cite{ikanat3}, 
and S. Pukas for his helpful suggestions.

%%%%%%%%%%% Insert your bibliography below %%%%%%%%%%

\end{document}